\documentclass [10pt]{article}
\usepackage{amsmath}
\usepackage{epsfig}

\begin{document}
\baselineskip=12pt \hsize=340pt \vsize=490pt

\def\b{\bar}
\def\d{\partial}
\def\D{\Delta}
\def\cD{{\cal D}}
\def\cK{{\cal K}}
\def\f{\varphi}
\def\g{\gamma}
\def\G{\Gamma}
\def\l{\lambda}
\def\L{\Lambda}
\def\M{{\Cal M}}
\def\m{\mu}
\def\n{\nu}
\def\p{\psi}
\def\q{\b q}
\def\r{\rho}
\def\t{\tau}
\def\x{\phi}
\def\X{\~\xi}
\def\~{\widetilde}
\def\h{\eta}
\def\bZ{\bar Z}
\def\cY{\bar Y}
\def\bY3{\bar Y_{,3}}
\def\Y3{Y_{,3}}
\def\z{\zeta}
\def\Z{{\b\zeta}}
\def\Y{{\bar Y}}
\def\cZ{{\bar Z}}
\def\`{\dot}
\def\be{\begin{equation}}
\def\ee{\end{equation}}
\def\bea{\begin{eqnarray}}
\def\eea{\end{eqnarray}}
\def\half{\frac{1}{2}}
\def\fn{\footnote}
\def\bh{black hole \ }
\def\cL{{\cal L}}
\def\cH{{\cal H}}
\def\cF{{\cal F}}
\def\cP{{\cal P}}
\def\cM{{\cal M}}
\def\ik{ik}
\def\mn{{\mu\nu}}
\def\a{\alpha}

\title{Twistor-Beam Excitations of Black-Holes and Prequantum Kerr-Schild Geometry}

\author{Alexander Burinskii, \\
NSI, Russian Academy of Sciences,\\
B. Tulskaya 52,  Moscow 115191 Russia}

\maketitle
\begin{abstract}
Exact Kerr-Schild (KS) solutions for electromagnetic excitations
of black-holes, have the form of singular beams supported on
twistor lines of the KS geometry. These beams have a very strong
back-reaction on the metric and horizon and create a fluctuating
KS geometry occupying an intermediate position between the
classical and quantum gravities. We consider the Kerr theorem,
which determines the twistor structure of the KS geometry and the
corresponding holographic prequantum space-time adapted to
subsequent quantum treatment.
\end{abstract}



1. Black-holes (BHs) are convenient objects for studying the
problem of unifying  quantum theory and gravity. One obstacle in
this very important problem  is the inconsistency of forms of
representations: gravity requires an explicit field representation
in the configuration space, while quantum theory consistently uses
the momentum space and plane waves, which, strictly speaking, are
not defined in gravity. Twistors form a bridge between these
representations. Geometrically, a twistor is  a pair $(x^\m, \
\theta^\alpha),$ where $\theta^\alpha$ is a two-component spinor
adjoined to the point $x^\m \in M^4$ and fixing the light
direction (beam) $\sigma^\m_{\dot\alpha \alpha}
\bar\theta^{\dot\alpha} \theta ^\alpha $ corresponding to the
momentum $p^m $ of a massless particle.
 A plane wave in the momentum space may be described in
the twistor coordinates $ T^I =\{\theta^\alpha, \ \mu_{\dot
\alpha} \}, \quad \mu_{\dot \alpha} = x_\n \sigma^\n_{\dot\alpha
\alpha} \theta^\alpha ,$ as $\exp \{i x_\m \sigma^\m_{\dot\alpha
\alpha} \tilde\theta^{\dot\alpha} \theta ^\alpha \} .$ On the
other hand, the wave can also be transformed to the  twistor space
by a "twistor" Fourier transform \cite{Wit}(p.2.5): a
multiplication by $-\exp(i\tilde\theta^{\dot\alpha}\mu_{\dot
\alpha} )$ and a subsequent integration over $\tilde\theta$. The
result\fn{$\hat\phi(T^I) =\int \frac {d^2 \tilde\theta}{(2\pi)^2}
\exp \{i x_\m \sigma^\m_{\dot\alpha \alpha}
\tilde\theta^{\dot\alpha} \theta ^\alpha
\}\exp(-i\tilde\theta^{\dot\alpha}\mu_{\dot \alpha} ) =
\delta^2(\mu_{\dot \alpha} - x_\n \sigma^\n_{\dot\alpha \alpha}
\theta^\alpha )$} $\hat\phi(T^I)=\delta^2(\mu_{\dot \alpha} - x_\n
\sigma^\n_{\dot\alpha \alpha} \theta^\alpha )$ corresponds to a
singular beam in the direction $p^\m$ with the twistor coordinates
$T^I =\{\theta^\alpha, \ \mu_{\dot \alpha}\}.$ It is essential
here that the twistor describes a beam passing through the given
point $x^\m.$

Similar twistor-beams appear in exact solutions of the
Debney-Kerr-Schild (DKS) equations \cite{DKS} and of compatible
 electromagnetic and gravitational fields
\cite{BurA}. The KS geometry covers a wide  class of algebraically
special metrics for rotating and non-rotating BHs and cosmological
solutions. The appearance of twistor-beams in the KS solutions is
not accidental: the twistor structure underlies the KS space-time.
Studying exact KS solutions therefore indicates a new way for
unifying quantum theory and gravity.

 The  exact nonstationary KS solutions considered bellow describe
 electromagnetic excitations of a BH in the form of fluctuating
 twistor -beams and also their exact back-reaction on the BH
 metric and horizon,  \cite{BurA}  consistent with the
 Einstein equations
 $R_\mn - \frac 12 g_\mn R = 8\pi <T_\mn> $.
 This fluctuating KS geometry has a specific two-sheeted holographic
 structure adapted to quantum treatment, \cite{Gib,SHW}, and
 occupies an  intermediate position between the classical and quantum
 gravities.

 2. The KS solutions are based on the KS metric
\be g_\mn =\eta_\mn + 2H k_\m k_\n , \label{KS}\ee where
$\eta_\mn$ is the metric of the auxiliary Minkowski space-time
$M^4 $ and $k_\m$ is a field of light directions forming a
congruence of null lines in $M^4 ,$ the so-called principal null
congruence (PNC) $\cal K .$ The Kerr and Kerr-Newman BH solutions
belong to algebraically special class (type D) of solutions  that
have two different PNCs, and the metric may consequently be
represented in two different forms: $ g_\mn^\pm =\eta_\mn + 2H
k_\m^\pm k_\n^\pm , $ were $k^{\m\pm}(x), \ (x=x^\m \in M^4)$ are
two different vector fields tangent to the corresponding $\cal
K^{\pm} .$  The directions $k_\m^{\pm}$ are determined in the
light-cone coordinates $ \ u=(z-t)/\sqrt {2},\quad v=(z+t)/\sqrt
{2},\quad\zeta=(x+iy)/\sqrt {2},\quad\bar\zeta=(x-iy)/\sqrt {2},$
by the differential 1-form\fn{The field $k^\m(x)\equiv k^\m(Y(x))
$ is completed to the adapted to the Kerr congruence null tetrad
$e^a , \ a=1,2,3,4 \ $ and "tetrad" derivatives $ ,_a=\d _a =
e^\m_a \d_\m . $} \be k_\m^{(\pm)} dx^\m = P^{-1}(du +\Y^\pm d\z +
Y^\pm d\Z - Y^\pm \Y^{(\pm)} dv), \label{kpm}\ee and depend on the
complex coordinate on celestial sphere \be Y=e^{i\phi} \tan \frac
{\theta}{2}  \label{Y} .\ee Two solutions $Y^\pm(x)$ are
determined by the  Kerr Theorem
\cite{DKS,Pen,PenRin,KraSte,Multiks,BurTwi} discussed bellow.

\begin{figure}[ht]
\centerline{\epsfig{figure=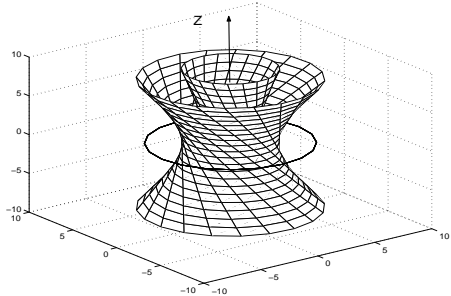,height=4cm,width=6cm}}
\caption{The Kerr singular ring and twistor rays of the Kerr
congruence. }
\end{figure}

The structure of Kerr congruence is shown in Fig.1. It has a
characteristic twisted form, which gives a complicated expression
for the Kerr metric despite the extremely simple general form of
the KS metric (\ref{KS}). The Kerr PNC consists of twistor light
beams covering the space-time twice: as the incoming PNC $k^{\m -}
\in \cal K^- $ which propagates toward the disk spanned by the
Kerr singular ring, and as the outgoing from disk PNC, $k^{\m +}
\in \cal K^+ ,$ located on the second sheet of the same spacetime
$M^4 .$ The metrics $g_\mn^\pm$ and  the corresponding
electromagnetic fields, being aligned with the beams of the PNC,
are different on the in- and out- sheets and should not be mixed
in KS solutions. This two-sheetedness is ignored in perturbative
approaches, which leads to solutions with drastic deviations from
the exact solutions.
  Typical {\it exact electromagnetic excitations on the KS
  background have the form of singular
beams propagating along the twistor lines of the PNC,} while
typical wave solutions in the perturbative approach have a smooth
angular dependence.

3.  The two-sheetedness was long an unsolvable problem with the
Kerr solution (see the references in \cite{BEHM3,BurSuper}).
Israel \cite{Isr} proposed truncating the second sheet and
replacing it with a rotating disk-like source covering the Kerr
singular ring and preventing the transition to the second sheet of
the metric. In an alternative variant, the Kerr singular ring
itself was considered as a closed "Alice" string forming a gate to
a mirror "Alice" world of  advanced vacuum fields \cite{DirKN}.
The new holographic approach unifies the two variants.  The source
of Kerr solution is formed by a domain wall (membrane) separating
the in- and out-sheets of the KS space, \cite{BurSuper}. The two
sheets of the KS geometry  and also the as the source-membrane are
needed for describing quantum fields in gravity. In particular,
according to Gibbons \cite{Gib}, the curved spacetime $\cal M$
should be separated into two causally ordered regions $\cal M_-$
and $\cal M_+$ associated with incoming and outgoing vacuum states
$|0_->$ and $|0_+> .$  If the source is absent, the basic KS
solutions can be extended analytically from the in-sheet to the
out-sheet, which effectively changes the sign of the frequency.
The presence of the source breaks this analyticity, separating the
retarded and advanced fields, and allows considering the BH
evaporation as a scattering of the in-vacuum on the membrane -
source. A similar prequantum BH space-time with separated in- and
out- sheets was also introduced  by 't Hooft et.al. in \cite{SHW}.

Two sheets of KS correspond to the 't Hooft holographic
correspondence in which the source of the Kerr BH forms a
holographically dual (to the bulk $\cal M_- \bigcup \cal M_+$)
boundary  separating the in- and out- regions.

The usual Penrose conformal diagram, containing the in- and
out-fields on the same $M^4 ,$ must be unfolded with a "splitting"
the KS two-sheetedness, as shown in Fig.2, demonstrating an
explicit realization  of the {\it holographic principle} in the KS
geometry. The twistor-beams of the Kerr congruence "create" the
Kerr source (as a holographic image) by light projection from the
past null infinity $I^- $ on the bulk of KS space-time. The BH
then appears as  a {\it holographic image} generated by the
initial data on the past infinity $I^- .$

\begin{figure}[ht]
\centerline{\epsfig{figure=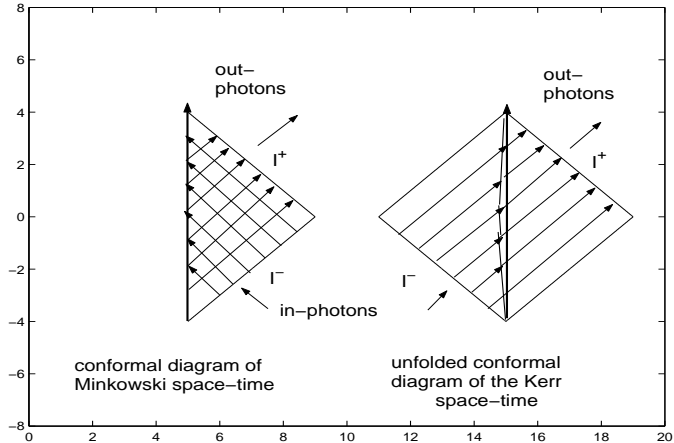,height=6cm,width=9cm}}
\caption{Penrose conformal diagrams for (a) the Minkowski
space-time and (b) the Kerr space-time: unfolding the auxiliary
$M^4$ space of the Kerr geometry into two sheets generates the
holographic structure of a prequantum BH space-time. }
\end{figure}

The KS light-like congruences are characterized by optical
parameters: divergence $\rho$, rotation (twist) $\omega$ and shift
$\sigma ,$  which describe the deformations of a two-dimensional
image projected by the beams on a distant two-dimensional screen.
The twist and divergence of the congruence lead to rotation and
scaling dilation  of the image, while the shift breaks the
conformal properties of the image, deforming angles. The
shear-free congruence of the Kerr solution $\sigma=0  $  retains
the conformal structure of the projected image under translations
of the screen along the beams of the congruence.\fn{The parameters
of dilatation $ \rho $ and twist $ \omega $ of the congruences are
combined in the KS formalism \cite{DKS} in one complex parameter $
Z= (\rho + i \omega), $ which can also be related with a complex
`distance' $\tilde r$, an affine parameter along the Kerr
congruence, $ \tilde r = P/Z,$ and $P$ is a conformal scale
factor.} The conformal structure and complex analyticity are
entered in the KS solutions via function $Y(x), \ x=x^\m \in M^4
,$ which is a conformal projection of celestial sphere $S^2$ (with
coordinates $(\phi,\theta)$) onto complex plane $Y\in C^1 .$
  Tetrad derivatives of the function $Y(x)$ determine principal
parameters of the KS holographic projection. In particular, {\it
the complex expansion} of congruence, $Z=\rho +i \omega ,$ is
determined from $Y(x)$ by the relation $ Z=Y,_1 \label{Y1} , $
{\it the geodesic condition} for the rays of congruence is $Y,_4
=0 ,$ and {\it the shear-free condition} is $ Y,_2 = 0.$
Therefore, the complex functions $ Y (x)$ obeying the conditions
\be Y,_2 = Y,_4 = 0 \label{Y24} \ee define the {\it shear-free and
geodesic congruences} possessing the conformal-analytic properties
of the KS holographic projection. All such congruences are
determined by  the Kerr Theorem which is formulated in terms of
the projective twistor coordinates\fn{The relation between twistor
coordinates $T^I=\{\theta^\alpha, \ \mu_{\dot \alpha} \}$ and
projective ones $Z^p$  is $T^I= \theta^1 ( 1,Z^p).$} \be Z^p = ( \
Y,\quad \z - Y v, \quad u + Y \Z ) .\ee

4. {\bf The Kerr Theorem,}
\cite{Pen,PenRin,KraSte,Multiks,BurTwi}. Any stationary geodesic
and shear-free null congruence in $M^4$ is generated by  solution
of the algebraic equation $ F(Z^p) = 0 , $ where $F$ is arbitrary
holomorphic function of the projective twistor coordinates $Z^p .$

For the Kerr congruence function $F$ is quadratic in $Y.$
 The performed in \cite{DKS} integration of the Einstein-Maxwell
equations showed that the function $F(Z^p)$ determines metric (\ref{KS}) of
the stationary KS solutions up to some arbitrary function $\psi(Y).$
Indeed, solution $Y(x)$ of the eq. $F=0$ determines congruence via
relation (\ref{kpm}) and also the function $H ,$ so far as it takes
the form
\noindent $ H = \frac 1 2 [ m (Z + \bar Z)P^{-1} - |\psi|^2 Z \bar
Z P^{-2}] , $ where $Z=Y,_1$ and $(Z/P)^{-1} = - dF/dY.$

In particular, for the Kerr geometry at rest $P=(1+Y\Y)/\sqrt{2},
\  \ P/Z = \tilde r = r+ia \cos \theta ,$ where $r, \ \theta$ are
oblate spheroidal coordinates, and \be H =\frac {mr - |\psi|^2/2}
{r^2+ a^2 \cos^2\theta} . \label{Hpsi} \ee

Electromagnetic (em) field is determined by vector potential
\be \alpha =\alpha _\m dx^\m \\
= -\frac 12 Re \ [(\frac \psi {r+ia \cos \theta}) e^3 + \chi d \Y
],   \label{alpha} \ee where $\chi = 2\int (1+Y\Y)^{-2} \psi dY  \ ,
$ which obeys
the alignment condition $ \alpha _\m k^\m=0 . $

\noindent
For the Kerr solution $m$ is mass of BH and $\psi =0 .$ The
Kerr-Newman solution has $\psi=q $ -- the charge parameter.
However, for the general stationary KS solutions {\it $\psi$
may be arbitrary holomorphic function of the complex variable $Y ,$}
 \cite{DKS}. On the other hand, any
nonconstant holomorphic functions on sphere acquire at least one
pole. A single pole at $Y=Y_i ,$  $\psi_i(Y) = q_i/(Y-Y_i)$
produces a singular em beam supported by the twistor of the Kerr PNC
with in angular direction $ Y_i=e^{i\phi_i} \tan \frac
{\theta_i}{2} \label{Yi} .$  In the same time, the function
$\psi(Y)$ acts immediately on the function $H$ which determines
the metric and the position of the horizon.
 The given in \cite{BEHM2} analysis showed
 that such em beams have very strong back reaction to metric
 and deform topologically the horizon, forming the holes in it, which
allows matter to escape interior of BH. The exact KS solutions with
several poles, $ \psi (Y) = \sum _i \frac {q_i} {Y-Y_i}, $  have
several twistor-beams with angular directions $Y_i=e^{i\phi_i} \tan
\frac {\theta_i}{2},$ leading to the horizon with many holes. In
the far zone the twistor-beams tend to the known exact singular
pp-wave solutions \cite{KraSte}.

5. The stationary KS twistor-beam solutions may be generalized to
time-dependent wave pulses, \cite{BurA}. Since the horizon is
extra sensitive to electromagnetic excitations, it may also be
sensitive to the vacuum electromagnetic field,  and the vacuum
fluctuations  may produce the beam pulses and a fine-grained
structure of fluctuating microholes in the horizon. It will allow
radiation to escape interior of the BH, as it is depicted on Fig.3,
leading to a semiclassical mechanism of BH evaporation.
\begin{figure}[ht]
\centerline{\epsfig{figure=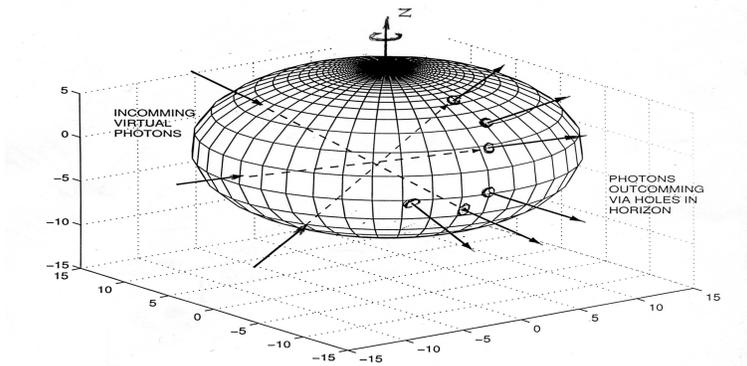,height=6cm,width=10cm}}
\vspace*{-7mm} \caption{Excitation of a rotating black-hole by the
field of virtual photons forming a set of fluctuating micro-holes
in its horizon.}
\end{figure}
Twistor-beam pulses have to depend on a retarded time $\t $ and
obey to the non-stationary Debney-Kerr-Schild (DKS) equations.
It was shown in \cite{DKS} that in this case, besides the function $\psi(Y,\t),$ the
exact KS solutions acquire extra radiative term $\gamma(Y,\t)Z .$
\fn{Tetrad components of the corresponding null electromagnetic
radiation have the form, \cite{DKS}, $F^\mn = Re \cF _{31}
e^{3\m}\wedge e^{1\n},$ where $ \cF _{31}=\gamma Z - (AZ),_1 \ .$
} The long-term efforts to integrating the time-dependent DKS equations,
\cite{BurA}, have led to consistent solutions with fluctuating
twistor-beam pulses. It was explicitly shown that any time-dependent
KS solution develops into the beam pulses breaking the topology and stability of the
BH horizon. Namely, interaction of a
black-hole with em field contains two different components:
a) the determined by function $\psi(Y,\t) $ singular twistor-beam pulses
which deform metric and topology of the horizon, but do not contribute
to energy flow; and b) a regular component  $\gamma_{reg}(Y,\t) $ which
determines em radiation and evaporation of the black-hole.

 The obtained solutions describe excitations of the
em twistor-beams on the Kerr-Schild background and consistent back reaction
of the beams to metric and  black-hole horizon. The holographic
space-time forms a fluctuating  pre-geometry which reflects dynamics of the
singular twistor-beam pulses. This pre-geometry
is classical, but it is twosheeted and has a fine-grained structure of fluctuating
twistor-beams which have to be still regularized to get the usual smooth classical
space-time.
In this sense it takes an intermediate position between the classical and
quantum gravity.
Note also, that the use of the Kerr theorem with the functions $F(Y)$ of
 higher degrees in $Y$ leads to multi-particle KS solutions
\cite{Multiks} which generate complicate networks of the twistor-beams.

\end{document}